\documentclass[twocolumn,showpacs,preprintnumbers,amsmath,amssymb]{revtex4}

\usepackage[dvips]{graphicx}
\usepackage{dcolumn}
\usepackage{bm}
\usepackage{color}

\newcommand{\BEQ}{\begin{equation}}     
\newcommand{\BEA}{\begin{eqnarray}}
\newcommand{\BD}{\begin{displaymath}}
\newcommand{\EEQ}{\end{equation}}       
\newcommand{\EEA}{\end{eqnarray}}
\newcommand{\ED}{\end{displaymath}}
 
\newcommand{\matz}[4]             
     {\mbox{${\begin{array}{cc} #1 & #2 \\ #3 & #4 \end{array}}$}}

\begin{document}


\title{
An alternative order-parameter for non-equilibrium generalized spin models on honeycomb lattices
}

\author{Francisco Sastre}
\email{sastre@fisica.ugto.mx}
\affiliation{%
Departamento de Ingenier\'ia F\'isica,\ Divisi\'on de Ciencias e Ingenier\'ias,
Campus Le\'on de la Universidad de Guanajuato,
AP E-143, CP 37150, Le\'on, M\'exico
}

\author{Malte Henkel}
\email{malte.henkel@univ-lorraine.fr}
\affiliation{%
Groupe de Physique Statistique, Institut Jean Lamour (CNRS UMR 7198), Universit\'e de Lorraine Nancy,
BP 70239, F -- 54506 Vand{\oe}uvre-l\`es-Nancy, France,
}

\date{\today}

\begin{abstract}
An alternative definition for the order-parameter is proposed, for a family of non-equilibrium
spin models with up-down symmetry on honeycomb lattices, and which depends on 
two parameters. In contrast to the usual definition, our proposal takes into account that each site of the lattice
can be associated with a local temperature which  depends on the local environment of each site. 
Using the generalised  voter motel as a test case, we analyse the phase diagram and the critical exponents in the
stationary state and compare the results  of the standard order-parameter with  the ones following from our new proposal, on the
honeycomb lattice. The stationary phase transition is in the Ising universality class. 
Finite-size corrections are also studied and the Wegner exponent is estimated as $\omega=1.06(9)$.
 

\keywords{Critical phenomena, voter model, critical exponents, Ising universality}
\end{abstract}

\pacs{05.20.-y, 05.70.Ln, 64.60.Cn, 05.50.+q}

\maketitle

\section{Introduction}

For equilibrium systems, the universality hypothesis allows to cast all 
critical systems in universality classes, of which the Ising model universality class is the best-known example. 
The concept of universality can be extended to non-equilibrium critical systems \cite{Henkel08,Tauber14}. 
In particular, a widely accepted conjecture states that non-equilibrium models with up-down symmetry and
spin-flip dynamics fall in the universality class of the Ising model~\cite{Grinstein1985}. 
A family of {\em generalized spin models} (GSM) that do not satisfy the detailed-balance
condition and which present a non-equilibrium steady-state, was proposed by Oliveira {\em et al.}~\cite{Oliveira1993,Tome2014}.  
The collective behaviour of the ``spins" shares many aspects with the well-established theory of non-equilibrium phase transitions and
results from simulations can be analysed similarly~\cite{Henkel08}. 
In these GSM, the system evolves following a competing dynamics induced by heat baths at two different temperatures
(on two-dimensional square lattices)~\cite{Droz1989,Tamayo1994,Drouffe1999} and hence have a {\em non}-equilibrium stationary state. 
In the original version of the GSM \cite{Oliveira1993}, 
each lattice site is occupied by a spin, $\sigma_i$, that interacts with its nearest neighbours. The system evolves in
the following way: during an elementary time step, a spin $\sigma_i = \pm 1$ on the lattice is randomly selected, and 
flipped with a probability given by
    \begin{equation}  \label{transition}
        p=\frac{1}{2}[1-\sigma_i f(H_i)],
    \end{equation}
where $H_i$ is the local field produced by the nearest neighbours to the site $i$ and $f(H_i)$ is a local function bounded by 
$|f(H_i)|\le 1$. On a square lattice, one  habitually uses one of two possible sets of parameters, such that 
    \begin{equation} \label{flipsquare}
    f(H_i) = \left\{ \begin{array}{l} f(2)=-f(-2)=x=\tanh(2\beta_2) \\
                                      f(0)=0 \\
                                      f(4)=-f(-4)=y=\tanh(4\beta_4)
                   \end{array} \right.
    \end{equation}
The dynamics is described by two parameters: either the pair ($x,~y$) which act analogously to a noise in the system, 
or else by a pair of effective inverse
temperatures ($\beta_2,~\beta_4$)~\cite{Oliveira1993,Drouffe1999}. 
In the second case, to each site one associates a ``temperature" that depends on its
instant local environment. This locally fluctuating ``temperature'' should affect the model's macroscopic behaviour.
Several known models are known special cases of the dynamics eqs.~(\ref{transition},\ref{flipsquare}): 
the {\em majority voter model} (MVM) corresponds to $y=x$ or $\beta_2=2\beta_4$;
the {\em Glauber-Ising model} (GIM) corresponds to $y=2x/(1+x^2)$ or $\beta_2=\beta_4$.
Numerical simulations confirm that these models, on a square lattice, belong to the 
Ising model universality class~\cite{Oliveira1992,Oliveira1993,Kwak2007,Wu2010}.
In  this work, we propose a new order-parameter with the following features: 
(i) it must take into account that the local variable has extra degrees of freedom because there is
not just one heat bath involved, and (ii) it must recover the standard Ising model when the heat baths are at the same temperature.
Additionally, we can introduce in this way a new model of out of equilibrium mixed-spin models, 
similar to the ferrimagnetic models (see \cite{Zukovic2015} and references therein).

This work is organised as follows: in section~\ref{modelo} we described how to implement the new order parameter on the honeycomb lattice. 
In section~\ref{FSS}, the finite-size
scaling method used to analyse its
stationary state is outlined. In section~\ref{resultados}, the results of the Monte Carlo simulation for a particular case, 
the equivalent MVM, are reported
and the critical parameters are extracted. 
We conclude in section~\ref{conclusiones}. 

\section{Model}  \label{modelo}

\begin{figure}[h!]
    \begin{center}
     \includegraphics[width=5.0cm,clip]{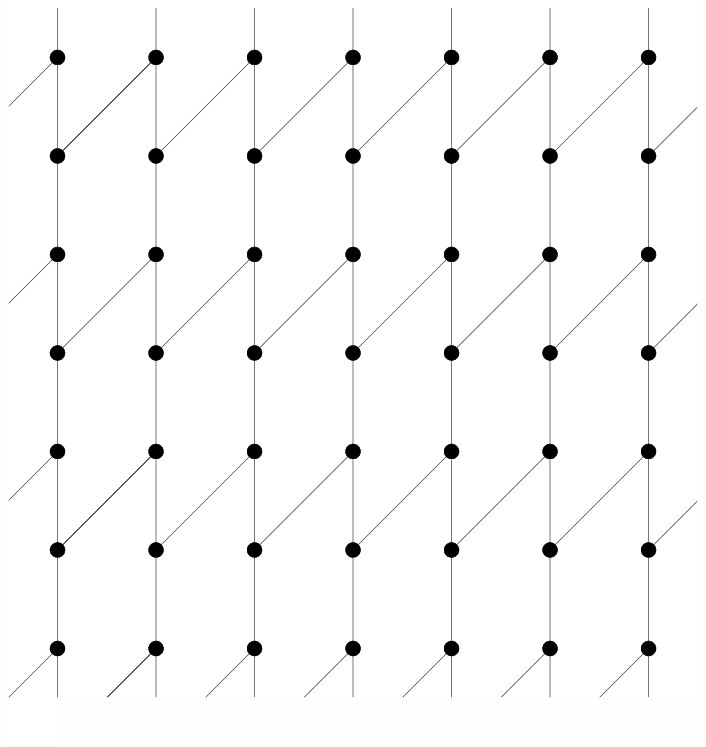}
     \caption[fig1]{\label{panal}
     Honeycomb lattice used for the simulations, with this geometry we use use skew boundary conditions at the horizontal boundary.
     }
    \end{center}
\end{figure}

In analogy with simple Ising magnets the paramagnetic-ferromagnetic phase transition can be measured with
the standard order-parameter, on a square lattice $\Lambda\subset\mathbb{Z}^2$, with $N=L^2=|\Lambda|$ sites
    \begin{equation} \label{4}
    \langle m\rangle =  \frac{1}{N} \left\langle\left|\sum_{i\in\Lambda} \sigma_i\right|\right\rangle,
    \end{equation}
In this work, we propose an alternative definition for an order parameter, as follows
    \begin{equation} \label{5}
    \langle \mu \rangle = \left\langle \left(\sum_{i\in\Lambda} \beta_{H_i}\right)^{-1} \left|\sum_{i\in\Lambda} \beta_{H_i} \sigma_i\right|\right\rangle.
    \end{equation}
such that the value of the inverse temperature $\beta_{H_i}$ is selected on each site, depending on the local field $H_i$, 
according to eq.~(\ref{flipsquare}). 
Certainly, when $\beta_2=\beta_4$, one should recover $\langle \mu \rangle = \langle m\rangle$. 
However, on lattices where each site has an even number of nearest neighbours, the order parameter $\langle\mu\rangle$ is not uniquely
defined, since for configurations with a local field $H_i=0$ at the site $i$, a further un-specified parameter $\beta_0$ must be introduced. 
A work-around is to consider lattices where sites have an odd number of nearest neighbours, such as the honeycomb lattice (see Fig.~\ref{panal}). 
For the honeycomb lattice, the available values for the local field are $H_i=\pm 1,~\pm 3$ and the extra parameter $\beta_0$ is no longer needed.

On the honeycomb lattice, eq.~(\ref{flipsquare}) is replaced by
    \begin{equation} \label{fliphoneycomb}
    f(H_i) = \left\{ \begin{array}{l} f(1)=-f(-1)=x=\tanh(\beta_1) \\
                                      f(3)=-f(-3)=y=\tanh(3\beta_3)
                   \end{array} \right.
    \end{equation}
Again, we recover some known models: 
the MVM corresponds now to $x=y$ and $\beta_1=3\beta_3$ and the GIM corresponds to 
$y=(3x+x^3)/(1+3x^2)$ or $\beta_1=\beta_3$.
Analogously we can define the susceptibility for the new order parameter as 
     \begin{equation}
     \chi = N x \{\langle \mu^2\rangle-\langle \mu\rangle^2\}.
     \label{susceptibility}
     \end{equation}

\begin{figure}[h!]
    \begin{center}
     \includegraphics[width=8.5cm,clip]{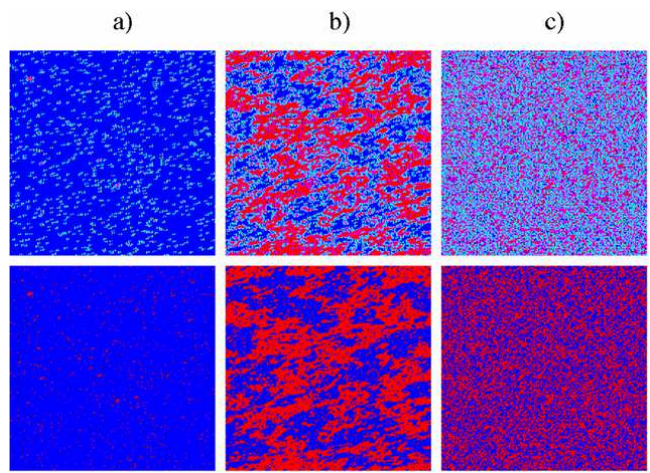}
     \caption[fig2]{\label{snapshots}
     (Colour online) Snapshots of the honeycomb lattice of size $L=200$ along the line $y=x$, for a) $x=0.95$ b) $x=0.8720$ and c) $x=0.20$.
     The upper row corresponds to the new order parameter $\langle\mu\rangle$, see (\ref{5}).
     Sites with a local spin $\sigma_i\beta_{H_i}=[+\beta_3, ~+\beta_1,~-\beta_1,~-\beta_3]$ are 
     represented with colours [blue, cyan, magenta, red], respectively. 
     The lower row corresponds to the standard order parameter $\langle m\rangle$, see (\ref{4}). 
     Sites with spin $\sigma_i=[+1,~-1]$ are represented by colours [blue,red], respectively. 
     The order-parameter values are 
     a) $\langle m\rangle \simeq 0.943,~\langle\mu\rangle\simeq 0.928$,
     b) $\langle m\rangle \simeq 0.052,~\langle\mu\rangle\simeq 0.040$ and
     c) $\langle m\rangle \simeq 0.009,~\langle\mu\rangle\simeq 0.007$.
    }
    \end{center}
    \end{figure}

For a qualitative illustration of the difference between the two order parameters, in
Fig.~\ref{snapshots} we present snapshots along the line $x=y$ for three different values of $x$ for a lattice of size $L=200$.
The left column a) corresponds to the ordered phase, the central column b) to the critical point and the right column c) to the
disordered phase. 
While the standard order parameter $\langle m\rangle$ permits to distinguish between the ordered and disordered phases, 
our new proposal $\langle\mu\rangle$ clearly hints at further hidden structures. For example, in the ordered phase, 
most sites have a local temperature $\beta_3$, whereas in the disordered phase, most sites have a local temperature $\beta_1$. 

    \begin{figure}[h!]
    \begin{center}
     \includegraphics[width=8.5cm,clip]{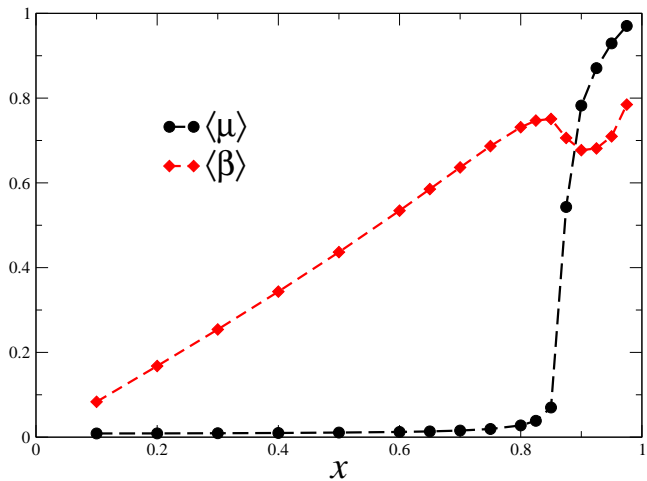}
     \caption[fig2bis]{\label{variacao_x}
     (Colour online) Dependence of the order-parameter $\langle\mu\rangle$  and
      of the average inverse temperature $\langle\beta\rangle$ on the coupling $x$ in the MVM, on the honeycomb lattice with $L=100$. 
    }
    \end{center}
    \end{figure}

In figure~\ref{variacao_x}, we illustrate the dependence of the averaged order-parameter $\langle\mu\rangle$ on the coupling $x$ in the MVM.
Clearly, the existence of a second-order phase transition is signalled, very analogous to what one has found for the usual order-parameter 
$\langle m\rangle$ \cite{Oliveira1992,Oliveira1993}. 
On the other hand, considering the average inverse temperature $\langle\beta\rangle$ hints at additional structure not captured by $\langle m\rangle$.
Curiously, there is a cusp in $\langle\beta\rangle$ which occurs very closely to the location $x_c$ of the critical point. While the
increase of $\langle\beta\rangle$ with $x$ in the disordered phase should mainly reflect the dependence of $\beta_1$ on $x$, 
the cusp could be related to the increase of sites with an inverse temperature $\beta_3=\frac{1}{3}\beta_1$, as is also suggested by the
snapshots in figure~\ref{snapshots}.

\begin{figure}[h!]
    \begin{center}
     \includegraphics[width=7.0cm,clip]{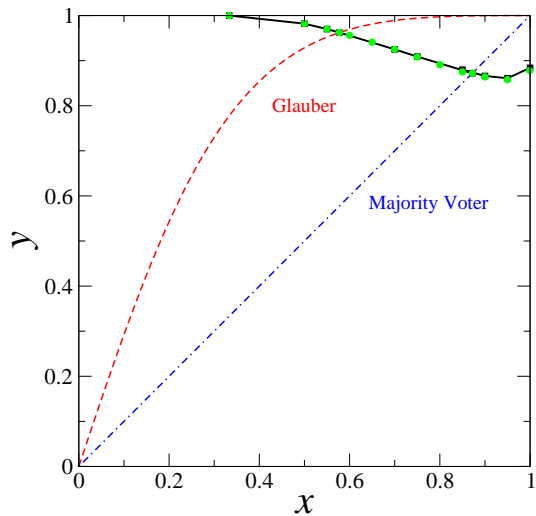}
     \caption[fig3]{\label{diagrama}
     (Colour online) Phase diagram in the $(x,y)$ plane. The ordered phase is in the right upper zone and 
     is separated from the disordered phase by the critical line, black
solid line for $\langle m\rangle$ and green circles for $\langle\mu\rangle$. 
The red dashed line correspond to the Ising model and the point-dashed blue line correspond to
the MVM.
    }
    \end{center}
    \end{figure}

Since both the MVM and the GIM present a continuous phase-transition, it is natural to assume that 
a critical line should exist in the $x-y$ plane. In analogy to the square 
lattice case, this line starts at the voter critical point, $(x,y)=(1/3,1)$ for the honeycomb lattice, 
and ends at the extremal value $(x,y)\approx (1,0.88)$. 
In order to sketch the critical line, we carried out rough simulations with small lattice sizes, 
$L=24, 28, 32$ and $36$, at different fixed $x$ values.
In order to estimate the critical points, we used the standard method of the crossing point of the fourth-order Binder cumulant \cite{Binder1981}
    \begin{equation}
    U^{(4)}=1-\frac{\langle \mu^4\rangle}{3\langle \mu^2\rangle^2}.
    \label{cumulant4}
    \end{equation}
For both order parameters (\ref{4},\ref{5})
simulations were carried out by starting with a random configuration of spins, 
and letting the system evolve according to the dynamics given by eqs.~(\ref{transition},\ref{fliphoneycomb}).
In figure~\ref{diagrama}, we show the phase diagram as estimated in the plane $(x,y)$ for the 
honeycomb lattice. Skew boundary conditions in the horizontal direction and 
periodic boundary conditions in the vertical direction were used, see Figure~\ref{panal}.
In all cases, the uncertainties in the critical values are around $10^{-3}$.
Clearly, we see from Figure~\ref{diagrama} that the critical boundaries estimated from both order-parameters are compatible 
(see also table~\ref{tab1}).
\begin{table}
\caption{\label{tab1} 
Rough estimates of the critical points as found from the order parameters $m$ and $\mu$, respectively, for the honeycomb lattice. 
We also included the points $(x,y)=(0.57735027,0.96225045)$, common to both, $(x,y)=(0.8721,0.8721)$ for $m$~\cite{AcunaLara2014}, 
and $(x,y)=(0.87195,0.87195)$ for $\mu$, as evaluated in next section.
}
\begin{center}
\begin{tabular}{ccc}  \hline\hline
$x$   & $y_c~(m)$ & $y_c~(\mu)$ \\ \hline
0.500 & 0.982     & 0.982 \\
0.550 & 0.970     & 0.970 \\
0.600 &  $-$      & 0.956 \\
0.650 &  $-$      & 0.941 \\
0.700 & 0.925     & 0.925 \\
0.750 & 0.909     & 0.908 \\
0.800 &  $-$      & 0.891 \\
0.850 & 0.880     & 0.875 \\
0.900 & 0.866     & 0.864 \\
0.950 & 0.861     & 0.858 \\
0.999 &  $-$      & 0.879 \\
1.000 & 0.884     &  $-$ \\
\hline\hline
\end{tabular}\end{center}
\end{table}

\section{Finite-size scaling technique} \label{FSS}

We shall use the method proposed in Ref.~\cite{Perez2005}, 
where three different cumulants are used for the
evaluation of the critical point: (i) the fourth-order or Binder cumulant, Eq.~(\ref{cumulant4}),
(ii) the third-order cumulant (where $\langle \mu^3\rangle$ is defined analogously to eq.~(\ref{5}))
    \begin{equation}
    U^{(3)}=1-\frac{\langle \mu^3\rangle}{2\langle \mu\rangle\langle \mu^2\rangle},
    \label{cumulant3}
    \end{equation}
and (iii) the second-order cumulant
    \begin{equation}
    U^{(2)}=1-\frac{2\langle \mu^2\rangle}{\pi\langle \mu\rangle^2}.
    \label{cumulant2}
    \end{equation}
The scaling forms for the thermodynamic observables, in the stationary state, 
and together with the leading finite-size correction exponent $\omega$ (or Wegner's exponent), are given by
    \begin{eqnarray}
    \mu(\epsilon,L) &\approx&  L^{-\beta/\nu}(\hat{M}(\epsilon L^{1/\nu})+
          L^{-\omega} \hat{\hat{M}}(\epsilon L)), \label{scaling3a}\\
    \chi(\epsilon,L) &\approx&  L^{\gamma/\nu}(\hat{\chi}(\epsilon 
          L^{1/\nu})+L^{-\omega} \hat{\hat{\chi}}(\epsilon L)),\\
    U^{(p)}(\epsilon,L) &\approx& \hat{U}^{(p)}(\epsilon L^{1/\nu})+
          L^{-\omega} \hat{\hat{U}}^{[p]}(\epsilon L). \label{scaling3}
    \end{eqnarray}
where $\epsilon=x-x_c$ is the distance from criticality, $p=2$, $3$ or $4$.
The parameters $\beta$, $\gamma$ and $\nu$ are the critical exponents for the infinite system, see \cite{Henkel08} for details.  

In principle, the critical point $x_c$ is found from the crossing points in the cumulants $U^{(p)}$. 
A precise estimation of $x_c$ is achieved by taking into account the crossing points for different cumulants $U^{(p)}$ and $U^{(q)}$ with
$p\ne q$ arise for different values of $L$.  
The values of $x$, where the cumulant curves $U^{(p)}(x)$ for two different linear sizes 
$L_i$ and $L_j$ intercept are denoted as $x^{(p)}_{ij}$.
We expand Eq.~(\ref{scaling3}) around $\epsilon=0$ to obtain
    \begin{equation}
    U^{(p)}\approx U^{(p)}_\infty+\bar{U}^{(p)} \epsilon L^{1/\nu} + 
        \bar{\bar{U}}^{(p)} L^{-\omega}+{\rm O}(\epsilon^2,\epsilon L^{-\omega}),
    \label{crossing1}
    \end{equation}
where $U^{(p)}_\infty$ are universal quantities, but $\bar{U}^{(p)}$ and 
$\bar{\bar{U}}^{(p)}$ are non-universal. The value of $\epsilon$ where 
the cumulant curves $U^{(p)}$ for two different linear sizes $L_i$ and 
$L_j$ intercept is denoted as $\epsilon^{(p)}_{i,j}$. At this crossing 
point the following relation must be satisfied:
     \begin{equation}
     L_i^{1/\nu}\epsilon^{(p)}_{ij}+B^{(p)} L_i^{-\omega} =
     L_j^{1/\nu}\epsilon^{(p)}_{ij}+B^{(p)} L_j^{-\omega}.
     \end{equation}
Here $B^{(p)}:=\bar{\bar{U}}^{(p)}/\bar{U}^{(p)}$. Combining for different cumulants ($q\ne p$) we get
    \begin{equation}
    \frac{x^{(p)}_{ij}+x^{(q)}_{ij}}{2}=x_{c}-(x_{ij}^{(p)}-x_{ij}^{(q)})A_{pq},
    \label{linealcrit}
    \end{equation}
where $A_{pq}=(B^{(p)}+B^{(q)})/[2(B^{(p)}-B^{(q)})]$ and is non-universal (see Refs.~\cite{Perez2005,AcunaLara2012} for additional details).
Equation~(\ref{linealcrit}) is a linear equation 
that makes no reference to $\nu$ or $\omega$ and requires as inputs only
the numerically measurable crossing couplings $x_{i,j}^p$. The intercept with the ordinate gives the critical point location.

\section{Results} \label{resultados}

For the determination of the critical point and the critical exponents for the MVM, we performed simulations on lattices with linear sizes
$L=24$, 32, 40, 48, 60, 76 and $96$, following the procedure used for the evaluation of the critical line of section~\ref{modelo}.
We let the system evolve during a transient time, that varied from $4\times10^5$ Monte Carlo time steps (MCTS) for $L=24$ to $1.5\times10^6$ 
MCTS for $L=96$. Averages of the observables were taken over $1\times 10^6$ MCTS for $L=24$ and up to $1.5\times 10^7$ MCTS for $L=96$. 
Additionally, for each value of $x$ and $L$, we performed 300 (bigger lattices) to 500 (smaller lattices) 
independent runs, in order to improve the statistics. 

    \begin{figure}[h!]
    \begin{center}
    \includegraphics[width=7.0cm,clip]{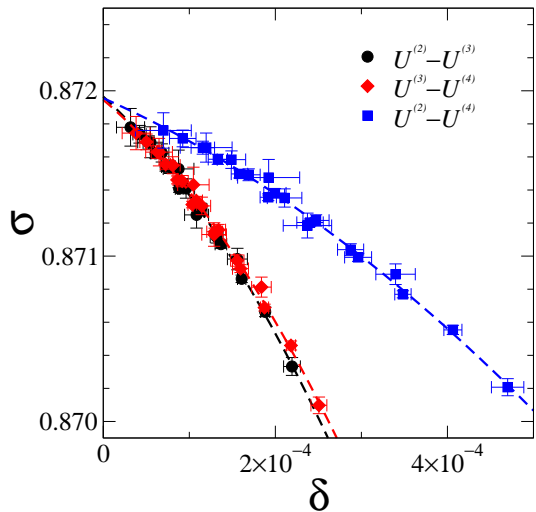}
    \caption[fig4]{\label{criticalpoint} (Colour online) Evaluation of the critical 
    point $x_c$.
    The points represent the numerical data obtained from third-order polynomial fits and the 
    dashed lines are second-order polynomial fits of Eq. (\ref{linealcrit}). Smaller $\delta$-values correspond
    to larger system sizes. 
    }
    \end{center}
    \end{figure}

For the evaluation of the critical points, we used a third-order polynomial fit 
for the cumulant curves. Recalling eq.~(\ref{linealcrit}), the estimation of the
critical point is shown in Figure~\ref{criticalpoint}, where we plot
the variable $\sigma :=(x_{ij}^{(4)}+x_{ij}^{(2)})/2$ over against the variable $\delta:=x_{ij}^{(4)}-x_{ij}^{(2)}$. 
We observe that the curves $\sigma(\delta)$ are not linear as expected from Eq.~(\ref{linealcrit}), 
this means that the finite-size effects are significant in this case
and the curvature is due to the neglected higher order terms in Eq.~(\ref{scaling3}).
When we compare the data for the crossing of the $U^{(2)}-U^{(4)}$ 
curves with the previous reported data for the standard order parameter given by Eq.~(\ref{4})
from Ref.~\cite{AcunaLara2014}, the range in the differences $\delta$ is almost two times larger 
with this new order parameter (see Fig. 2c in Ref.~\cite{AcunaLara2014}) 
and that the smallest difference
is around $7\times10^{-5}$ (corresponding to the crossing between $L=96$ and $L=76$).
When we compare with results for the antiferromagnetic MVM on honeycomb lattices 
(Figure 4b on reference~\cite{Sastre2015}) we observe that the scaling effect, like the range in the
differences $\delta$ and the departure from the linear behavior of $\sigma(\delta)$, are more notorious with the new order parameter.
With the second order polynomial fits of Eq. (\ref{linealcrit}) 
we obtain the result for the critical point of 
\BEQ
x_c=0.87195(22)
\EEQ
where the number in brackets give the
estimated uncertainty in the last given digit(s). This results is in good agreement with the reported value 
for the critical point of the standard order parameter for the
ferromagnetic and antiferromagnetic MVM on honeycomb lattice~\cite{AcunaLara2014,Sastre2015}.

Once that we have the critical point, we can also analyse eventual finite-size corrections, which are described in terms
of Wegner's correction-to-scaling exponent $\omega$, which was already defined in eqs.~(\ref{scaling3a}-\ref{scaling3}). 
We evaluate the Wegner exponent $\omega$ and the universal quantities 
$U_{\infty}^{(p)}$ by using a non linear fit with~(\ref{scaling3}) and $\epsilon=0$.
    \begin{figure}[h!]
    \begin{center}
    \includegraphics[width=7.0cm,clip]{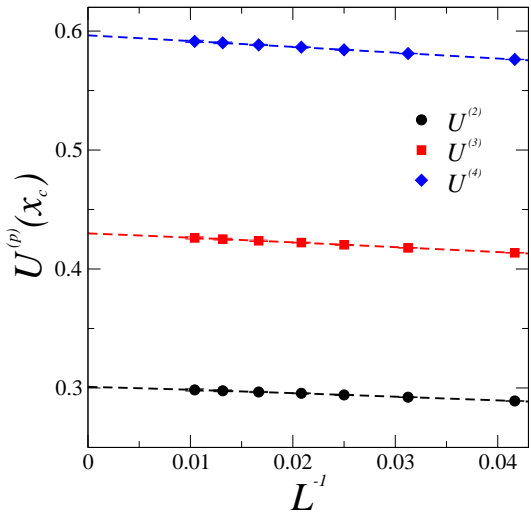}
    \caption[fig5]{\label{bindergraf} (Colour online)
    Cumulant values at the critical point as function of the inverse linear size.
    The dashed lines are non linear fits of Equation~(\ref{scaling3}).
    }
    \end{center}
    \end{figure}
Again, we can observe that the scaling effects are more pronounced here, compared to the antiferromagnetic case 
with the standard order parameter on the same lattice (see Figure 8b in~\cite{Sastre2015}).
Our estimated value for the Wegner exponent is 
\BEQ
\omega=1.06(9)
\EEQ 
The usual prediction for $\omega$, based on $2D$ conformal invariance \cite{Henkel99,Caselle2002,Izmailian2011} gives $\omega=2$
for the Ising model, on a square lattice with periodic boundary  conditions. Early suggestions that $\omega$ might  be as small as $4/3$ have been disproved numerically, in favour of $\omega=2$ \cite{Blote1998}. However, in certain cases the corresponding
amplitude may vanish and this leads to an effective value $\omega=4$, as seen for the Ising model on honeycomb and 
triangular lattices \cite{Queiroz2000}. For open boundary conditions or Brascamp-Kunz boundary conditions, one rather finds $\omega=1$
\cite{Salas2002,Queiroz2011,Kenna2002}. Finally, on the triangular lattice, effective values in the range $1.2 \lesssim \omega_{\rm eff}\lesssim 2$
were reported \cite{Guo2005}.

The values for the cumulants are $U_{\infty}^{(2)}=0.301(6)$, $U_{\infty}^{(3)}=0.430(7)$ and $U_{\infty}^{(4)}=0.597(7)$; 
they are all compatible with the reported values
for the antiferromagnetic MVM on honeycomb lattices with the same boundary conditions~\cite{Sastre2015}.

\begin{figure}[h!]
    \begin{center}
    \includegraphics[width=7.0cm,clip]{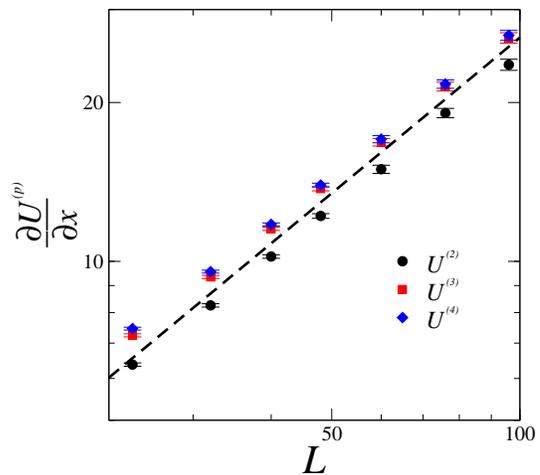}
    \caption[fig6]{\label{nuexponent} (Colour online)
    Log-log plot of the derivatives of the cumulants $U^{(2)}$, $U^{(3)}$ 
    and $U^{(4)}$ with respect to $x$, taken {\em at} the critical point $x=x_c$. 
    The dashed line show the expected power-law behaviour in the $L\to\infty$ limit for $1/\nu=0.98$.
    }
    \end{center}
    \end{figure}

\begin{figure}[h!]
    \begin{center}
     \includegraphics[width=7.0cm,clip]{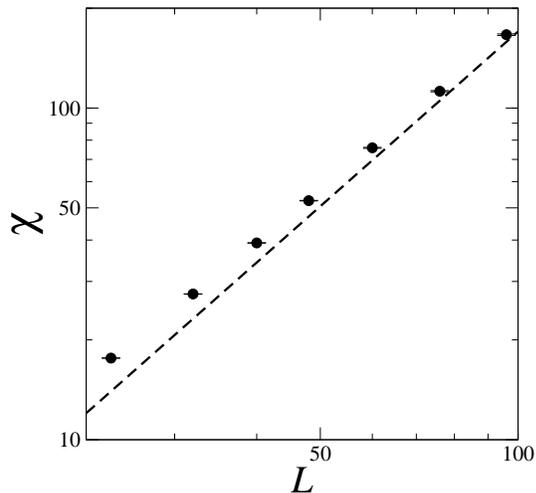}
     \caption[fig7]{\label{gammaexponent}
     Log-log plot of the susceptibility at the critical point.
     The dashed line show the expected power-law behaviour in the $L\to\infty$ limit for $\gamma/\nu=1.75$.
    }
    \end{center}
    \end{figure}

\begin{figure}[h!]
    \begin{center}
    \includegraphics[width=7.0cm,clip]{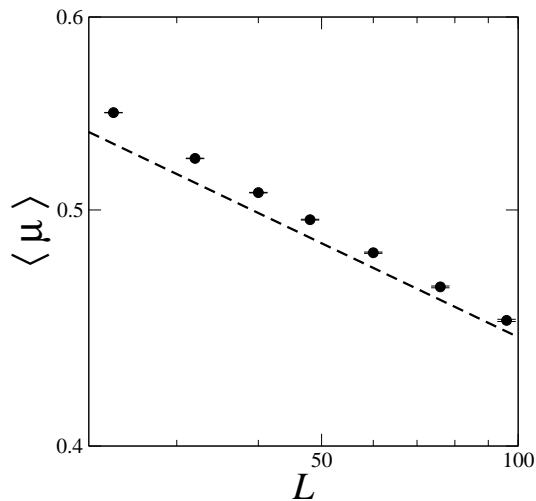}
    \caption[fig8]{\label{betaexponent}
    Log-log plot of the order parameter at the critical point.
    The dashed line show the expected power-law behaviour in the $L\to\infty$ limit for $\beta/\nu=0.128$.
    }
    \end{center}
    \end{figure}

The critical exponents can be evaluated, by using Eqs.~(\ref{scaling3}), at the critical point $\epsilon=0$. 
One expects the following finite-size scaling behaviour 
    \begin{eqnarray}
    m(L) &\sim& L^{-\beta/\nu}(1+a_m L^{-\omega}), \label{beta}\\
    \chi(L) &\sim& L^{\gamma/\nu}(1+a_{\chi} L^{-\omega}), \label{gamma}
    \end{eqnarray}
and
    \begin{equation}
    \frac{\partial U^{(p)}}{\partial x}\Bigl|_{x=x_c} 
        \sim L^{1/\nu}(1+a_{U^{(p)}}L^{-\omega}),
    \label{nu}
    \end{equation}
where the correction-to-scaling exponent, with the value  $\omega=1.06$, must be included, 
since the correction-to-scaling effects are important in this case. The parameters $a_{\alpha}$ are non-universal.
In Fig.~\ref{nuexponent}, we show the derivatives of the cumulants at the critical point. From the finite-size scaling law~(\ref{nu}),
we obtain the following results: 
$1/\nu=0.97(4)$, 0.99(4) and 0.99(3) from $\partial U^{(2)}/\partial x$, $\partial U^{(3)}/\partial x$ and $\partial U^{(4)}/\partial x$, 
respectively. After combining our results we finally obtain $1/\nu=0.98(4)$, in good agreement with the
result for the two-dimensional Ising model.
The evaluation of $\gamma/\nu$ is shown in Fig.~\ref{gammaexponent}, with the relation (\ref{gamma}) 
we obtain $\gamma/\nu=1.75(2)$.
We present in Fig.~\ref{betaexponent} the evaluation of $\beta/\nu$, with Eq.~(\ref{beta}) we obtain $\beta/\nu=0.128(5)$.

%
%

All reported numerical estimates are very close to the exactly known values of the two-dimensional Ising ferromagnet, 
e.g. \cite[appendix A]{Henkel08}. 

\section{Conclusions} \label{conclusiones}

We have introduced a new definition of an order-parameter for a family of generalized 
spin models in honeycomb lattices. This definition (\ref{5}) of the new order parameter $\mu$ can be extended to other lattice types with an odd
number of nearest neighbours. 
We studied through intensive Monte Carlo simulations the stationary state 
for a particular case that corresponds to the Majority voter model.
We have found that the phase-diagram for this order-parameter in the 
two-parameter space $(x,y)$ is equivalent to the phase diagram for the standard order
parameter, in both ferromagnetic and antiferromagnetic versions of the MVM. Furthermore, the estimated critical exponents
\BEQ
\frac{1}{\nu} = 0.98(4),\quad \frac{\gamma}{\nu}=1.75(2),\quad \frac{\beta}{\nu}=0.128(5)
\EEQ
of the stationary state are, as expected, 
compatible to the ones of the $2D$ Ising model universality class. 
One important difference with respect to the previous numerical studies 
concerns the the correction-to-scaling effects. Their analysis permits to 
estimate the Wegner exponent $\omega$. Our result $\omega_{\rm MVM}=1.06(6)$ is not compatible with the conventional value $\omega_{\rm Ising}=2$
of the two-dimensional Ising model. Such small values of $\omega$ have usually been reported on certain non-periodic lattices in the
$2D$ Ising model \cite{Salas2002,Queiroz2000,Kenna2002}  while here, it arises as a result of the chosen dynamics. 
The only other known value for the MVM is for the
three-dimensional cubic lattice~\cite{AcunaLara2012}; in this case the value of $\omega$ is compatible with the value of the Ising model. 
Further simulations should be performed in order to test further whether there is a new correction exponent for this model.

Finally, the use of $\mu$ might offer a  finer view on the mechanism of a phase transition, see figures~\ref{snapshots} and~\ref{variacao_x}, 
and probably can be extended beyond the Ising model. 
We  hope to return elsewhere to further exploration
of new insights, especially in the context of non-equilibrium phase transitions. 
 
\section*{Acknowledgments}

FS thanks the Groupe de Physique Statistique \`a l'Universit\'e de Lorraine Nancy for warm hospitality. This work 
was partly supported by the Coll\`ege Doctoral franco-allemand Nancy-Leipzig-Coventry
({\it `Syst\`emes complexes \`a l'\'equilibre et hors \'equilibre'}) of UFA-DFH.

\newpage
\bibliography{bibliografia}

\end{document}